\documentclass[12pt]{article}

\usepackage{latexsym}

 \font\tenmsbm=msbm10 scaled 1200
\font\sevenmsbm=msbm9
\newfam\msbmfam
\textfont\msbmfam=\tenmsbm \scriptfont\msbmfam=\sevenmsbm

\def\msbm{\fam\msbmfam\tenmsbm}

\makeatletter \@addtoreset{equation}{section} \makeatother
\renewcommand{\theequation}{\thesection.\arabic{equation}}

\newcounter{parentequation}
\newenvironment{subequations}{%
  \refstepcounter{equation}%
  \begingroup
\let\protect\noexpand
  \edef\@tempa{\def\noexpand\theparentequation{\theequation}}%
  \expandafter
  \endgroup\@tempa
  \setcounter{parentequation}{\value{equation}}%
  \setcounter{equation}{0}%
  \def\theequation{\theparentequation\alph{equation}}%
  \ignorespaces
}{%
  \setcounter{equation}{\value{parentequation}}%
}
\newcommand{\eqn}[1]{(\ref{#1})}
\def\IC{\relax\,\hbox{$\inbar\kern-.3em{\rm C}$}}

 \def\cB{{\cal B}}

\def\cN{{\cal N}} 
 
\def\cR{{\cal R}} 
 
\def\beq{\begin{equation}}
\def\eeq{\end{equation}}
\def\bea{\begin{eqnarray}}
\def\eea{\end{eqnarray}}
\def\bet{\begin{tabular}}
\def\eet{\end{tabular}}
\def\bes{\begin{subequations}\bea}
\def\ees{\eea\end{subequations}}

\def\ZZ{\hbox{\msbm Z}}

\begin{document}

\begin{titlepage}

\begin{flushright}
HU-EP 01/30\\
hep-th/0107264
\end{flushright}

\vspace{2cm}

\baselineskip 8 mm

\begin{center}

{\LARGE \bf Type IIB supergravity compactified on a Calabi--Yau 
manifold with $H$--fluxes}

\baselineskip 5 mm

\vspace{2cm}

Gianguido Dall'Agata$^{\sharp}$

\vspace{1cm}

{$\sharp$ \it Institut f\"ur Physik, Humboldt Universit\"at \\
Invalidenstra\ss{}e 110, 10115 Berlin, Germany\\
{\tt dallagat@physik.hu-berlin.de}}
\end{center}

\vspace{2cm}

\begin{abstract}
We discuss the compactification of type $IIB$ supergravity on a 
Calabi--Yau manifold in the presence of both $RR$ and $NS$ fluxes for 
the three--form fields.
We obtain the classical potential both by direct compactification and 
by using the techniques of ${\cal N}=2$ gauged supergravity in 
four--dimensions.
We briefly discuss the properties of such potential and compare the 
result with previous derivations.
\end{abstract}

\vskip 20mm


\end{titlepage}

\newpage

\baselineskip 6 mm

\section{Introduction}

It is a known fact that the compactification of type $IIB$ 
supergravity on a six--dimensional Calabi--Yau manifold gives rise to 
an effective four--dimensional theory which is given by ${\cal N}=2$ 
supergravity coupled to $h_{(2,1)}$ vector multiplets and 
$h_{(1,1)}+1$ hypermultiplets.
The addition of non--trivial fluxes for the Neveu--Schwarz and Ramond 
three forms make the universal hypermultiplet charged under the 
axionic shift symmetries of the moduli--space.
As a consequence, the scalar sector develops a 
non--trivial potential whose critical points determine new vacua of 
the theory, possibly breaking supersymmetry to ${\cal N} = 1$ or ${\cal 
N} = 0$.

The phenomenon of partial supersymmetry breaking occurs very rarely
for ${\cal N} = 2$ theories in $D = 4$ \cite{susy0,susy1,susy2} and therefore it is
very interesting to see if there is any chance to obtain it in this
framework.
Moreover, this scenario could provide a mechanism to construct ${\cal 
N} = 1$ gauge theories in the limit of decoupling of gravity.

As one could expect, this kind of theories have already
been considered in the past.
The general features of the compactifications without fluxes have been
discussed in \cite{CDF}, whereas the details have been
worked--out in \cite{Halle}.

For what concerns the theories in the presence of fluxes, the vacua of 
the derived potential have been studied in \cite{Michelson}, claiming 
that one can have only ${\cal N} =2$ or ${\cal N} = 0$ supersymmetry 
preserved.
In \cite{TV} it is further argued that one can produce ${\cal N} = 1$ 
potentials considering non--compact Calabi--Yau manifolds and taking 
a certain decompactification limit that makes some 
degrees of freedom non--dynamical.
In particular, it was derived the expression of the final ${\cal N} = 1$ 
superpotential $W$ in terms of the fluxes, according with the general 
analysis of \cite{Gukov}.
In \cite{Mayr,Berlino} this same $W$ was related to the four--dimensional 
quaternionic prepotentials and it was performed the analysis of the 
supersymmetric vacua of such theory for singular Calabi--Yau manifolds.

In this paper we address once more the problem of the derivation of 
the four--dimensional theory and the study of its vacua.
We will derive the classical potential of the four--dimensional 
effective theory both by direct compactification and 
by using the techniques of ${\cal N}=2$ gauged supergravity.
As expected,  the two results agree. 
This fact, that could seem trivial, was instead a controversial 
point of previous derivations.
In particular, it was known that the potentials derived by Michelson \cite{Michelson} 
(using the ${\cal N}=2$ gauged supergravity techniques) and 
by Taylor and Vafa \cite{TV} (using direct compactification) 
did not agree.
For instance the potential presented in \cite{TV} is positive semi--definite, 
whereas the one in \cite{Michelson} is not.
Moreover, as noted in \cite{Mayr}, the potential of
\cite{TV} is also $SL(2,\ZZ)$ invariant and this feature is not present 
in the one of \cite{Michelson}.

We also show that the resulting potential has a run--away behaviour.
This implies that no critical points at all are allowed in the 
classical region of validity of the scalar--manifold variables.
In particular, to obtain critical points, the volume of the Calabi--Yau 
is driven to infinity. 
We don't discuss here the possibility of using singular manifolds 
along the lines of \cite{TV,Mayr,Berlino}, 
which we leave for the future.

We will also briefly discuss some interesting properties of such 
potential, with a special attention to the possibility of writing it 
in the ${\cal N}=1$ form presented in \cite{TV}.
We will show that this {\it cannot} be done for the generic case, but that only 
{\it a truncation} of some of the hyper--scalars can lead to such a form.
Although this cannot be done retaining an ${\cal N} = 2$ theory, it 
could happen that for some consistent reduction to ${\cal N} = 1$ 
this possibility appears.

We will also discuss the supersymmetry laws of the Fermi fields, 
showing that indeed one cannot find ${\cal N} = 1$ configurations 
unless the hyperinos are discarded.
This leaves open the possibility to obtain a consistent ${\cal N} = 
1$ theory in the limit where some of the moduli are freezed.

\section{Compactification of type $IIB$ supergravity}

Let us start computing what is the effect of turning on a v.e.v. for the three 
form field strength in type IIB supergravity when we compactify the 
theory down to four dimensions on a Calabi--Yau manifold.

For the sake of semplicity, we will consider here a manifold with 
arbitrary $h_{(2,1)}$ but $h_{(1,1)}=1$, which is the minimum 
required by a Calabi--Yau.
This implies that the number of hypermultiplet is fixed to 
two and that they parametrize the quaternionic manifold given 
by $G_{2(2)}/SO(4)$ \cite{FerraraSabharwal}.

On the Calabi--Yau manifold $\cal M$, we take a canonical homology 
basis for $H_3$ given by $A^\Lambda, B_\Lambda$ ($\Lambda = 0,1,\ldots, b_{2,1}$) and 
let $(\alpha_\Lambda, \beta^\Lambda)$ be the dual cohomology basis.
Completeness of this basis implies that 
\begin{equation}
    \int_{A^\Lambda} \alpha_\Pi = \int_{\cal M}   \alpha_{\Pi}  \beta^\Lambda  = \delta^\Lambda_{\Pi} ,
\end{equation}
as well as the converse
\begin{equation}
    \int_{B_{\Pi}} \beta^\Lambda = \int_{\cal M}  \beta^\Lambda  \alpha_{\Pi}   = -\delta^\Lambda_{\Pi} .
\end{equation}

Since we are going to discuss the effect of turning on $H$--fluxes, we 
will give an expectation value to the Ramond and Neveu--Schwarz three--forms 
as
\begin{equation}
\langle \, H^1 \rangle = e^1_{\Lambda} \beta^\Lambda, \quad  \langle\, 
H^2\rangle = e^2_{\Lambda} \beta^\Lambda.
    \label{H}
\end{equation}

This, of course, is not the most general form of fluxes one can turn 
on, but in what follows we will limit the analysis only to electric charges.
We choose  to impose this restriction, because we want to make a 
comparison with results that can be obtained from the standard  four--dimensional 
gauged supergravity and therefore the electric charges are
the only ones which can receive a correct treatment in such framework 
\cite{4d}.
Anyway it is not difficult to believe that the results could be  
extended to the presence of magnetic charges, as well.
It is known that these charges can be obtained by a rotation in 
the symplectic base used for describing the special K\"ahler geometry 
of the vector scalar manifold.

We expect that the new contributions to the ten--dimensional theory coming 
from the above fluxes \eqn{H} 
are given by new terms or modifications to the kinetic terms of the three 
and five forms and to the Wess--Zumino term.
These terms will produce a four--dimensional scalar potential as well 
as new interactions between the scalar and vector fields of the theory.
In particular they will make charged some of the scalars under two abelian 
gauge symmetries.

Since we are mainly interested in the final scalar potential, in the following 
we will consider only the bosonic sector of such theory.
This is indeed enough to read directly the result for such potential, 
as well as to read the quantities needed to identify the  
four--dimensional gauged theory which reproduces such result.

\ 

Our starting point will be the covariant action for the ten--dimensional 
theory of type $IIB$ supergravity of \cite{IIBbose,IIB}.
In this action, the problem of obtaining the self--duality constraint of 
the $RR$ five--form without compromising Lorentz invariance is 
solved by using the PST formalism \cite{PST}.
This implies the addition of a scalar field $a(x)$, singlet under 
supersymmetry,  which is a pure gauge under additional symmetries 
that the action now exhibits.

Of course, when we compactify this action to four dimensions, this extra field 
should not contribute and so we just set it to zero, discarding then 
also all the terms which contain it. 
Doing this, the relevant part of the starting action is given 
by \cite{IIB}
\begin{equation}
 \int d^{10}x \sqrt{-g} \, R \,+  \frac{1}{4}\int F_5 \star F_5 + 2 \int \bar{F}_{3} \star F_3 + 2 
 \int \bar{F}_1 \star F_1 + \frac{\rm i}{2} \int (B \bar H - \bar B H) 
 F_5
    \label{eq:action}
\end{equation}
where the definition of the various forms is
\begin{equation}
\begin{array}{rclrcl}
F_1 & = & \bar U d V - V d \bar U,
&
F_3 & = & \bar U H + V \bar H,  \\
F_5 & = & d A_4 + {\rm i} (B \bar H - \bar B H),
&
H &=& dB,
\end{array}
\end{equation}
and $\star$ denotes the Hodge--duality operation.
The $(U,V)$ scalars are the dilaton/axion of the ten dimensional 
theory  parametrizing the $SU(1,1)/U(1)$ 
manifold.
This implies that they must satisfy the constraint $|U|^{2} - |V|^2 = 
1$.
At the same time they define the $U(1)$ connection
$
    Q  = -\frac{\rm i}{2}(\bar U d V - V d \bar U).
$

\ 

From this action we now want to compute the scalar potential of the 
four--dimensional theory, consequence of the introduction of the 
\eqn{H} fluxes.

To do so and also make contact with previous work, 
we make the following field redefinitions
\begin{equation}
\begin{array}{rclcrcl}
U & = &   \displaystyle  \frac{1}{\sqrt{1 - \psi \bar \psi}},  & \quad&
V & = &   \displaystyle  \frac{\psi}{\sqrt{1 - \psi \bar \psi}},
\end{array}
\end{equation}
with
\begin{equation}
\psi = \frac{1-\eta}{1+\eta}, \ \hbox{ and } \ \eta = \alpha + i 
\beta, \quad (\alpha = e^{-\varphi_{10}}).
\end{equation}

We can also decompose $H$ in its real and imaginary part $H = 
H_1 + {\rm i} H_2$, such that we can rewrite the kinetic term for the 
three--forms as
\begin{equation}
2 \bar F_3 \star F_3 = \frac{4}{\eta + \bar \eta} [ (H_1 - {\rm i} \, 
\bar \eta \, H_2) \star (H_1 + {\rm i} \, \eta \,  H_2 ) ].
    \label{cambio}
\end{equation}

Defining now
\begin{equation}
    e_{\Lambda} \equiv  e^1_{\Lambda}  + {\rm i} \eta e^2_{\Lambda}  
\end{equation}
we can compute the potential following the lines of \cite{TV}.
Indeed, we can write \eqn{cambio} as
\begin{equation}
2\; (Re \, \eta)^{-1} \int \bar e_{\Lambda} \beta^\Lambda \wedge \star 
    (e_{\Sigma} \beta^\Sigma )
    \label{kin}
\end{equation}
and make use of the duality relations between the Hodge--forms 
for a Calabi--Yau threefold \cite{Suzuki}. 

If indeed we call ${\cal R} \equiv Re {\cal N}$, ${\cal I} \equiv Im {\cal 
N}$, where $\cN_{\Lambda \Sigma}$ is the matrix of the kinetic term for the vector 
fields in $D = 4$, the Hodge duality of the forms reads
\begin{eqnarray}
\star \alpha_\Lambda & = & ({\cal RI}^{-1})_\Lambda{}^\Sigma 
\alpha_\Sigma - ({\cal I} + {\cal RI}^{-1}{\cal R})_{\Lambda\Sigma} 
\beta^\Sigma,
    \label{eq:al}  \\
\star \beta^\Lambda & = & {\cal I}^{-1 \,\Lambda \Sigma} 
\alpha_\Sigma - ({\cal I}^{-1}{\cal R})^{\Lambda}{}_\Sigma \beta^\Sigma.
    \label{eq:bet}
\end{eqnarray}
Using \eqn{eq:al},\eqn{eq:bet} and the relation coming from the 
special K\"ahler geometry \cite{CDF}
\begin{equation}
U^{\Lambda\Sigma} \equiv f^\Lambda_i g^{i\bar \jmath} f_{\bar 
\jmath}^\Sigma=  -\frac12 {\cal I}^{-1 \Lambda \Sigma}  - \bar L^\Lambda L^\Sigma
\end{equation}
it is straightforward to show that 
\eqn{kin} becomes 
\begin{equation}
\frac{8}{\eta + \bar \eta} \;  e_{\Lambda} \bar{e}_{\Sigma}\left(\bar L^\Lambda 
 L^\Sigma + U^{\Lambda \Sigma}\right).
 \label{potz}
\end{equation}
 
This (besides normalizations) is the same result obtained in \cite{TV}.
Anyway, we will see in a moment that this is not the final result for
the potential of the theory, since, to go in the Einstein frame and
obtain the correct normalization for the kinetic terms, we still need
to rescale the metric.
This additional factor will reveal crucial to obtain an agreement 
with the gauged supergravity result.

\ 

We now show that the above potential \eqn{potz} does not receive 
further contributions from the other terms in the action and also fix 
the amount of the rescaling needed to go to the four--dimensional 
Einstein frame.
We will also compute the various kinetic terms for the scalars and 
vectors and their couplings to determine the form of the 
four--dimensional supergravity action to which we want to compare our 
results.
To simplify the reading of this technical part, we divide the 
various sectors of the compactification according to the expansion of 
the various fields into different harmonic forms.
Moreover, we will mainly quote the results, detailing only the 
derivation of some very important structures.

First we will compute the contributions of the five--form field
(through its kinetic term and the Wess--Zumino term) to the kinetic
terms of the vectors and the couplings of some scalars with the
vectors through the $H^3$ sector.
Then we will analyze the $H^2$ and $H^0$ sectors computing the kinetic 
terms of the various four--dimensional scalars.
This will fix completely the four--dimensional theory and determine 
the gauging one needs to perform.

Let us start with the $H^3$ sector.
The zero modes of the five--form on the Calabi--Yau manifold, now that 
we have included also the three--form fluxes, are given by
\begin{equation}
F_5 = F^\Lambda \alpha_\Lambda - G_\Lambda \beta^\Lambda+ 2\, 
B^i  \,  e_\Lambda^j \, \beta^\Lambda \, \epsilon_{ij}.
    \label{eq:defF}
\end{equation}
This implies that the five--form kinetic term gives the following 
four--dimensional action
\begin{eqnarray}
&-& \frac{1}{4} \int \left\{F^\Lambda \star F^\Sigma ({\cal I} + {\cal RI}^{-1}{\cal 
R})_{\Lambda\Sigma} + 2 G_{\Lambda} \star F^{\Sigma} ({\cal 
RI}^{-1})_\Sigma{}^\Lambda - G_{\Lambda} \star G_\Sigma {\cal I}^{-1 
\Lambda \Sigma} \right. + \nonumber \\
&-& \left. 4 \left[F^\Lambda ({\cal I}^{-1} {\cal 
R})^\Sigma{}_\Lambda + G_\Lambda {\cal I}^{-1\,\Lambda \Sigma} \right] B^i 
e^j_{\Sigma} \epsilon_{ij}\right\}.
\end{eqnarray}
Although the starting action was formulated such as to produce the 
correct ten--dimensional equations of motion, including the 
self--duality constraint on the five--form, to compactify the theory 
we had to set the auxiliary scalar $a(x)$ to zero.
Unfortunately, this means that in doing so we lost the self--duality property.
To restore it, we add to the above action the Lagrange multiplier
\begin{equation}
    \frac{1}{2} G_{\Lambda} F^\Lambda.
\end{equation}
Integrating out $G_\Lambda$ one produces the constraint restoring 
self--duality (see also \cite{CDF}) 
\begin{equation}
G_\Lambda = {\cal I}_{\Lambda \Sigma} \star F^\Sigma + {\cal R}_{\Lambda 
\Sigma} F^\Sigma + 2 \, B^i  \,  e_\Lambda^j \, \epsilon_{ij}
    \label{eq:gL}
\end{equation}
and the kinetic five--form term reduces to
\begin{equation}
 \frac{1}{2} \left( {\cal R}_{\Lambda \Sigma } F^{\Lambda} 
F^\Sigma + {\cal I}_{\Lambda \Sigma } F^{\Lambda} 
\star F^\Sigma\right) +  B^i  \,  e_\Lambda^j \, \epsilon_{ij} \, F^\Lambda .
    \label{eq:kinF5}
\end{equation}
The first part is just the kinetic term of the vector fields of 
four--dimensional gauged supergravity \cite{4d}, whereas the 
last term reduces to a change of normalization of the reduction 
of the Wess--Zumino ten--dimensional term, which, after integrating by 
parts, we can write as
\begin{equation}
2 \int H^i \,  e_\Lambda^j \, \epsilon_{ij} \, A^\Lambda .
\label{norm}
\end{equation}
As one can understand the three--form field strength in four 
dimensions will be dualized to a scalar field strength and therefore 
the above coupling determines the scalars which are charged under the 
symmetries gauged by $A^\Lambda$. 

\ 

In a similar fashion, one can compute
the kinetic terms and the couplings for the scalar fields coming from 
the $H^2$ and $H^0$ sectors.
In doing this we use the simplificating assumption that for our 
Calabi--Yau manifold $h_{(1,1)}=1$ and therefore in the $H^2$ sector 
all the forms are proportional to the K\"ahler structure $J$.
This fixes also the duality relation of any $Y \in h_{(1,1)}$, which 
is given by
\begin{equation}
\star Y = \frac{3}{2} \frac{\int Y \wedge J \wedge J}{\int J \wedge J \wedge 
J} - Y \wedge J.
\end{equation}

The five--form $F_5$ and the three--forms $H^i$ will give rise to 
three scalar degrees of freedom when expanded in the $H^2$ sector:
\begin{eqnarray}
H^i &=& d a^i \, Y, \\
F_5 &=& d c \; Y \wedge Y + f_3 \, Y + 2 (a^i H^j +B^i d a^j) 
\epsilon_{ij} \, Y + 2 a^i \, da^j \epsilon_{ij} \, Y \wedge Y.
\end{eqnarray}

Once again, to implement the self--duality of the five--form in the 
resulting theory, one has to add to the standard action a multiplier 
between $f_3$ and  $d {\cal C} \equiv d c + 2 a^i \, d a^j \epsilon_{ij}$.
The integration of $f_3$ will leave with the kinetic terms for the 
physical $c$ and $a^i$ fields.

The fourth scalar in this sector comes from the Calabi--Yau metric 
${\rm i} g_{i\bar \jmath} = e^\sigma Y_{i\bar \jmath}$ and its kinetic term 
comes from the ten--dimensional Einstein term.

The four scalars of the universal hypermultiplet come instead from 
the $H^0$ sector.
For instance from the ten--dimensional  one--forms we have:
\begin{equation}
2 \int \bar F_1 \star F_1 = 2 \int \frac{ d  \bar \eta \star d  \eta}{(\eta + 
\bar \eta)^2} = \int \frac{1}{2 \alpha^2} \left( d  \alpha \star  d  
\alpha + d \beta \star  d \beta \right).
    \label{eq:f1kin}
\end{equation}
The other two come from the ten--dimensional three--forms properly dualized.
To perform this dualization one has to add further  Lagrange
multipliers $b^i$ through
\begin{equation}
- H^i db^j \epsilon_{ij}.
    \label{eq:ff}
\end{equation}

\ 

To obtain the effective four--dimensional theory, we still have to 
remember that the integration over the internal manifold gives a 
factor of $e^{3 \sigma}$ in front of the four--dimensional Einstein 
term.  
Therefore, to go back to the Einstein frame and to 
normalize the Einstein kinetic term as in \cite{4d}, 
we have to reabsorb it by 
a proper Weyl rescaling of the four--dimensional metric
$g_{\mu\nu} \to \frac12 e^{-3 \sigma} g_{\mu\nu}$.

Defining the covariant derivative $D b^i = d b^i + 2 e_\Sigma^i
A^\Sigma$ and, for ease of notation\footnote{The asymmetry between 
$\cB^1$ and $\cB^2$ is due to the integration by parts of the terms 
of the form $B^i da^j a^k da^l \epsilon_{ij} \epsilon_{kl}$. } 
$D {\cal B}^1 \equiv D b^1 - 24 a^1( 
d {\cal C} - 2 a^1 \, d a^2 )$, $D {\cal B}^2 \equiv D b^2 - 24 a^2( 
d {\cal C} - 4 a^1 \, d a^2 )$, one obtains as final action for the four--dimensional
scalars:
\begin{equation}
\begin{array}{rcl}
S_{scal} &=& \displaystyle \int_{4d} \left[ \frac{1}{4 \alpha^2} \left( d  \alpha \star  d  
\alpha + d \beta \star  d \beta \right) + 3 e^{-4 \sigma} \, d {\cal C} 
\star d {\cal C} + 3 d \sigma \star d \sigma +\right.\\
& & \\
&+& \displaystyle \frac{1}{16 \,\alpha}  e^{-6 
\sigma} \left(D\cB^1 - {\rm i} \bar \eta 
D\cB^2\right)\star \left(D\cB^1 +{\rm i}
\eta D\cB^2 \right) + \\
&& \\
&+& \left.\displaystyle \frac{3}{\alpha} e^{-2 \sigma} \left(da^1 - {\rm i} \bar \eta 
da^2\right)\star \left(da^1 + {\rm i} \eta da^2\right)\right].
\end{array}
\label{eq:Sscal}
\end{equation}

\ 

It is known that these scalars should parametrize a dual quaternionic 
manifold.
To identify the underlying quaternionic structure we now propose  
some identifications between the above scalars and the usual variables 
used to parametrize these manifolds \cite{FerraraSabharwal}.
This will become useful for the comparison with the gauged 
supergravity results, where the new coordinates are more natural.

Although in this computation we have chosen to limit 
ourselves to the analysis of the scalars 
describing the hyper sector of the moduli 
space of a Calabi--Yau with $h_{(1,1)} = 1$, which is the 
minimal set one can consider, we expect that the result will hold 
also in the generic case, provided one modifies the  
change of coordinates.

As follows from the the following identifications, the quaternionic
manifold described by the above scalars is the coset $G_{2,2}/SO(4)$
\cite{BC}
\begin{eqnarray}
z & = &  4 a^2 - 2\, {\rm i}\frac{e^{\sigma} }{\sqrt{\alpha}},
\label{eq:Z}  \\
S & = & e^{3 \sigma} \sqrt{\alpha} -  \frac{\rm i}{2} b^1 - 
\frac{1}{4} (C + \bar C) {\cal R}^{-1}(C + \bar C),
\label{eq:S}  \\
C_0 & = & - \beta e^{3 \sigma} \alpha^{-3/2} - \hbox{Re} \,z \, 
\hbox{Re}\, C_1+ 
\frac{\rm i}{2} b^2,
\label{eq:C0}  \\
C_1 &=& 3\frac{e^\sigma}{\sqrt{\alpha}} a^1 + 3 {\rm i} (c- 2 
a^1 a^2), \\
K & = & - \log \left( -\frac{\rm i}{2} (z - \bar z)^3\right),
\label{eq:K}  \\
\tilde{K} & =& - \log \left( S + \bar S + \frac{1}{2} (C + \bar C) {\cal 
R}^{-1}(C + \bar C)\right),
\label{eq:Kt}  
\end{eqnarray}
where
\begin{equation}
    \cN = \frac{\rm i}{8}  \left( \begin{array}{cc} \displaystyle-\frac14 (4 z^3 + 
    12 z^2 \bar z - 3 z \bar z^2 + 6 \bar z^3)& 3(z^2 + z 
    \bar z)\\ 3(z^2 + 
    z \bar z)& -9 z - 3 \bar z\end{array}\right), 
\end{equation}
and $\cR = (\cN + \bar \cN)/2$.
This implies that one can write the kinetic terms for the above scalars as
\begin{equation}
\int d^4 x \;\sqrt{-g}\; h_{uv}  \, D_{\mu} q^u D^\mu q^v = 
 \int \left( u \star \bar u + v \star \bar v + e \star \bar e + E 
\star \bar E \right),
    \label{eq:Squat}
\end{equation}
where \cite{FerraraSabharwal}
\begin{eqnarray}
u &  =  &  2 e^{(\tilde K + K)/2} Z \left( dC - \frac12 d {\cal N} 
R^{-1} (C +\bar C)\right),\\
v & = & e^{\tilde K} \left( d S + (C + \bar C) R^{-1}dC - \frac{1}{4} 
(C + \bar C) R^{-1} d {\cal N} R^{-1} (C +\bar C)\right),\\
e &=& P \, dZ,\\
E &=& e^{(\tilde K - K)/2} P \, N^{-1} \left( dC - \frac12 d {\cal N} 
R^{-1} (C +\bar C)\right),
\end{eqnarray}
are the general complex vielbeins, and in our case
\begin{eqnarray}
Z &=& \{1,z\},\\
P &=& \left\{ - \sqrt{3} \frac{z}{z-\bar z}, \frac{\sqrt{3}}{z-\bar 
    z}\right\}, \\
N &=& \frac14 \left( \begin{array}{cc} 2 {\rm i} (z^3 - \bar z^3) & -3 
{\rm i} (z^2 - \bar z^2)  \\ -3 {\rm i} (z^2 - \bar z^2)  & 6 {\rm i} (z - \bar z) 
\end{array}\right)
\end{eqnarray}
and the other quantities have been defined above.

We also see now that $b^1$ and $b^2$ correspond to -Im$S$ and Im$C_0$ 
respectively and the introduction of fluxes is producing a gauging of 
the Peccei--Quinn isometries $S \to S + {\rm i} a$ and $C_0 \to C_0 + {\rm i} a$.

\ 

We are now in position to express the potential of the theory in terms 
of this variables.
Although this will make the expression of the potential rather 
complicated, it will be useful in the comparison with the results 
coming from the four--dimensional gauged supergravity.

The potential is given by the term we obtained from the three--forms
\eqn{potz} rescaled with the same rescaling we used for
the scalar fields.  
The ten--dimensional dilaton can be written as
\begin{equation}
\eta = 4 e^{( K-\tilde K)/2} + {\rm i} \left[(C + \bar C) 
R^{-1}\right]^0,
\end{equation}
and the rescaling
\begin{equation}
\frac{e^{-6 \sigma}}{4} = 4 \, e^{( 3 \tilde K-K)/2},
\end{equation}
therefore the final expression is 
\begin{eqnarray}
{\cal V}&=& 4 \left(\bar L^\Lambda 
 L^\Sigma + U^{\Lambda \Sigma}\right) \left[   e^{2 \tilde{K}} \, 
 e_{\Lambda}^1 e_{\Sigma}^1 + 
 e^{2\tilde K}  \, \left( 16  \, e^{K-\tilde K} \,+ \left(\left[(C + \bar C) 
 \, R^{-1}\right]^0\right)^2  \right) \,  e_{\Lambda}^2 e_{\Sigma}^2 \right. \nonumber\\
 &-& \left.  \left[(C + \bar C) \, R^{-1}\right]^0 \,
 e^{2\tilde K}\;(e_{\Lambda}^1 e_{\Sigma}^2 + e_{\Lambda}^2 e_{\Sigma}^1)\right].
 \label{potzresc}
\end{eqnarray}

\section{The potential from gauged supergravity} 

Let us now compare this potential with the one that would be expected
from the gauging.

The Killing vectors corresponding to the shift symmetries in Im$S$ and Im$C_0$
\begin{equation}
k_\Lambda = -2 e^2_{\Lambda}\left( \begin{array}{c} i \\ -i 
\\0\\0\end{array}\right)+2 e^1_\Lambda 
\left(\begin{array}{c} 0\\0\\i \\ -i \end{array} \right). 
\label{killing}
\end{equation}
From these vectors one could compute the quaternionic prepotentials 
$P^x_{\Lambda}$ from their standard definition
\begin{equation}
i_\Lambda \Omega^x = \nabla P^x_\Lambda,
\end{equation}
where $\Omega^x$ is the $SU(2)$ curvature of the quaternionic manifold 
and $\nabla$ is the $SU(2)$ covariant derivative.
In our case these vectors correspond to Peccei--Quinn symmetries under 
which the Lie derivative of the $SU(2)$ connection vanishes and 
therefore the above expression can be simplified \cite{Michelson}. 
The prepotentials can be given by the 
contraction of this Killing vectors with the $SU(2)$ connection:
\begin{equation}
{\cal P}_\Lambda^x = \omega^x_u k^u_{\Lambda}.
\end{equation}
Since the potential which comes from gauging abelian 
isometries of the quaternionic manifold can be 
written as \cite{4d}:
\begin{equation}
{\cal V} = (U^{\Lambda \Sigma} - 3 \bar L^\Lambda L^\Sigma) 
P_\Lambda^x P_\Sigma^x + 4 \bar L^\Lambda L^\Sigma h_{uv} k^u_{\Lambda} 
k^v_{\Sigma},
\end{equation}
or, in terms of the prepotentials, as
\begin{equation}
{\cal V} = (U^{\Lambda \Sigma} - 3 \bar L^\Lambda L^\Sigma) 
P_\Lambda^x P_\Sigma^x + \frac13 \bar L^\Lambda L^\Sigma h^{uv} 
\nabla_u P_{\Lambda}^x \nabla_v P_{\Sigma}^x \, ,
\end{equation}
we just need the explicit expression of the connection to obtain it.

The quaternionic vielbeine which give the scalar action 
\eqn{eq:Squat} are 
\begin{equation}
    {\cal U}^{\alpha A} = \frac{1}{\sqrt{2}} \left( 
    \begin{array}{cccc} u & e & - \bar v & - \bar E \\ v & E & \bar 
    u & - \bar e \end{array}\right).
\end{equation}
From these we can define as usual the metric and the curvature
\begin{eqnarray}
h_{uv} &=& {\cal U}_u^{\alpha A} {\cal U}_{v}^{\beta B} \, 
C_{\alpha \beta} \epsilon_{AB}, \\ 
\Omega^x &=& {\rm i} C_{\alpha \beta} \, (\sigma^x)^C_A\epsilon_{CB} \, 
{\cal U}^{\alpha A} \wedge {\cal U}^{\beta B} = d \omega^x + 
\frac12 \epsilon^{xyz} \omega^y \omega^z,
\end{eqnarray}
and from this latter, we can derive the connection,
which is given by
\begin{eqnarray}
\omega^1 & = &  {\rm i} (u - \bar u),
    \label{eq:om1}  \\
\omega^2 & = &  (u + \bar u),
    \label{eq:om2}  \\
\omega^3 & = & \frac{\rm i}{2} (\bar v - v) + \frac{\rm i}{2} \frac{\bar Z N dZ - Z N d \bar 
Z}{\bar Z N Z}.
    \label{eq:om3}
\end{eqnarray}

Contracting now with the Killing vectors \eqn{killing} we obtain 
the prepotentials
\begin{equation}
{\cal P}_\Lambda = \left( \begin{array}{c} 8 e^{(\tilde K + K)/2} 
e_\Lambda^2 \\ 0 \\ 2 e^{\tilde K} e^1_\Lambda - 2 e^{\tilde K} [(C + 
\bar C) R^{-1}]^0 \, e_\Lambda^2\end{array}\right).
\end{equation}

For the case under consideration such  prepotentials satisfy the 
non--trivial relation  
\begin{equation}
 P_{\Lambda}^x P_\Sigma^x = h_{uv} k_\Lambda^u k_\Sigma^v.
    \label{equal}
\end{equation}
This implies that the final structure of the potential goes through a 
drastic simplification and is given by 
\begin{equation}
{\cal V} = (U^{\Lambda\Sigma} + \bar{L}^\Lambda L^\Sigma)\left( P_{\Lambda}^x 
P_{\Sigma}^x\right).
    \label{potfin}
\end{equation}
By a direct evaluation one can see that it is the same one as that presented in 
\eqn{potzresc}.

This result differs from the one presented in \cite{Michelson}, due to the 
relation \eqn{equal} that now is satisfied.
This happens because the truncation performed in \cite{Michelson}  
looses a contribution to the square of the Killing vectors $h_{uv} k^u 
k^v$ which comes from the $E$ complex  vielbein which was discarded.
If one instead first considers the full metric (which contains also the $E 
\otimes \bar E$ term) and then truncates to the subsector described by 
$C_1 = 0$, the result agrees with the one we presented.
Indeed, the above--mentioned term gives an additional 
contribution equal to 
\begin{equation}
e^{{\tilde K} - K}[P N^{-1}]^0 [\bar P \bar N^{-1}]^0.
\label{eq:add}
\end{equation}
Using the general property
\begin{equation}
P \cdot P^\dagger = - \frac{1}{\bar Z N Z}\left( N - \frac{(NZ)(\bar Z 
N)}{\bar Z N Z}\right),
\end{equation}
and the equality
\begin{equation}
N^{00} = - \frac{2}{\bar Z N Z},
\label{N00}
\end{equation}
which holds for the $G_{2(2)}/SO(4)$ coset, one obtains that
\eqn{eq:add} gives the additional contribution which is essential to
make the equation \eqn{equal} satisfied.

\section{Comments}

Inspecting the equation defining the potential
\begin{equation}
{\cal V} = (U^{\Lambda\Sigma} + \bar{L}^\Lambda L^\Sigma)\left( P_{\Lambda}^x 
P_{\Sigma}^x\right),
\label{potenzial}
\end{equation}
one can see that it is positive definite, i.e. ${\cal 
V}\geq 0$ and that it vanishes if and only if $\left( P_{\Lambda}^x 
P_{\Sigma}^x\right) = 0$.

It is easy to show that the requirement of stationary points under variation 
of the hypermultiplets implies that all the coupling constants be 
zero.
The potential has indeed a run--away behaviour in this sector as can 
be understood by expressing the above potential in terms of the 
four--dimensional dilaton $\phi_4 \sim e^{\tilde K}$.
Also, going back to the original ten--dimensional variables, one sees 
that this same potential is a function of the ten--dimensional 
dilaton/axion $\eta$ and the inverse volume of the Calabi--Yau $e^{-6 \sigma}$.
Variation under $\sigma$ shows a critical point for the 
decompactification limit $\sigma \to \infty$.

\ 

An interesting feature of this potential is that it can be written in 
a form which is very close to that of a pure ${\cal N} = 1$ 
supergravity theory.
If we use  the "${\cal N} = 1$ section"
\begin{equation}
L = e^{\frac{1}{2} {\cal K}} W,
\end{equation}
where we defined the superpotential \cite{Mayr,Berlino}
\begin{equation}
W = e^{- {\cal K}/2} L^\Lambda (P_\Lambda^3 \pm {\rm i} P_\Lambda^1), \quad {\cal K} = 
K_V -(K +\tilde K), 
\label{defW}
\end{equation}
the potential becomes
\begin{equation}
{\cal V} = -3 \bar L L + g^{i\bar \jmath} \nabla_i L \nabla_{\bar 
\jmath} \bar L +  h^{uv} \nabla_u L \nabla_v \bar L ,
\label{N1pot}
\end{equation}
with 
\begin{equation}
\nabla_{A} \equiv \partial_A + \frac12 \partial_A {\cal K} .
\end{equation}
Although the form of \eqn{N1pot} looks like that of a pure ${\cal N} = 
1$ potential, it is not yet so, since the quaternionic 
metric $h_{uv}$ cannot  in general be derived from the K\"ahler 
potentials $K + \tilde K$.
It has been shown \cite{FerraraSabharwal} that the necessary condition for such metric to be 
also K\"ahler is given by the holomorphicity of the matrix ${\cal 
N}_{\Lambda\Sigma}$.
This restricts the possible quaternionic manifolds to be 
$SU(2,n+1)/SU(2)\times SU(n+1) \times U(1)$,
which includes the truncation to the Universal 
hypermultiplet alone.
Unfortunately, as one can argue from the above example, 
these are not the manifolds which are chosen by the 
Calabi--Yau compactifications.
As for such manifolds the equation \eqn{N00} does not hold anymore, the
quaternionic prepotentials will not satisfy \eqn{equal} and this 
prevents us from the possibility of writing the potential in the 
\eqn{potenzial} form and as a further consequence as
\eqn{N1pot}.

The only chance to make of the \eqn{N1pot} potential a real 
${\cal N} = 1$ potential is to truncate the theory to the degrees 
of freedom given by the 
ten--dimensional dilaton/axion scalars $\eta$, the volume $\sigma$ 
and its axionic partner $c$.
This cannot be achieved by just fixing their value to a constant in 
the ${\cal N} = 2$ theory, as the obtained potential has no critical 
points at all.
It could anyway be that there is a consistent truncation of such 
theory to an ${\cal N} = 1$ one, where the above scalars become chiral 
multiplets\footnote{For the general conditions under which such a 
truncation is consistent we refer the reader to \cite{futuro,Louis}.
Related aspects were also presented in J. Louis' seminar at Humboldt 
Uni. in May 2001.}.
In performing this truncation, the form of the $ {\cal N} =1$ 
superpotential $W$ must be consistent with the reduction of the 
supersymmetry laws and this fixes it to be of the form given in 
\eqn{defW}.

The final theory will be then a real ${\cal N} = 1$ supergravity 
theory, whose potential form \eqn{N1pot} (with $h_{uv}$ now a K\"ahler 
metric) would then be justified.

A further consistency problem is given by the fact that $W$  is holomorphic in the 
coordinates of the vector scalar manifold, but it is not in general in terms of 
the hypers\footnote{We thank R. D'Auria and S. Ferrara for discussions
on this point.}.

\ 

One can also understand that the theory we are describing is a genuine 
${\cal N}=2$ theory analyzing the supersymmetry laws of the 
various Fermi fields.
The shifts of the gravitini, gaugini and hyperini are given by 
\begin{equation}
S_{AB} = - \frac{\rm i}{2} \left( \begin{array}{cc} -P^1 & P^3 \\ 
P^3 & P^1 \end{array}\right)_{AB},
\label{SAB}
\end{equation}
where now $P^x \equiv L^\Lambda P_\Lambda^x$, 
\begin{equation}
W^{iAB} =  {\rm i} g^{i\bar \jmath} \left( \begin{array}{cc} 
- {\nabla_{\bar \jmath} } \bar P^1 &  {\nabla_{\bar \jmath} } \bar P^3 \\ 
 {\nabla_{\bar \jmath} } \bar P^3 &  {\nabla_{\bar \jmath} } \bar P^1 
 \end{array}\right)^{AB},
\end{equation}
and
\begin{equation}
{\cal N}^{\alpha A} =  \frac{\rm i}{\sqrt{2}} \left( 
\begin{array}{cc} 
-\bar P^1 & \bar P^3\\
0 & i_{\bar k} E\\
 \bar P^3 &  \bar P^1 \\
 - i_{\bar k} \bar E & 0
\end{array}\right)^{\alpha A}.
\end{equation}

From the gravitini transformation law, one notice that if there is a 
vacuum such that 
\begin{equation}
    P^1 = \pm {\rm i} P^3 \qquad \Leftrightarrow 
    \quad W = 0
\label{condiz}
\end{equation}
the matrix \eqn{SAB} becomes degenerate and therefore one can preserve half
supersymmetry.

Due to the reality properties of the quaternionic prepotentials, this
same condition \eqn{condiz} implies that the $L^\Lambda$ sections
cannot be chosen to be all independent.
This, as a further consequence, implies that there is no holomorphic 
prepotential for the K\"ahler potential of the vector scalar manifold 
\cite{FerraraSabharwal}.

At the same time, one can see that the gaugini transformation law 
implies that also the derivatives must satisfy an analogous relation 
\begin{equation}
    \nabla_i P^1 = \pm \, {\rm i}\, \nabla_i P^3 \qquad \Leftrightarrow 
    \quad \nabla_i W = 0.
\end{equation}
This is still possible to satisfy again by a choice of section which 
is not linearly independent \cite{susy1,susy2}.

Then there are the hyperini transformations.
As one can see, the same condition \eqn{condiz}  applied to the ${\cal 
N}^{\alpha A}$ matrix implies that half of the hyperini supersymmetry 
laws can vanish, but still the other
half cannot.
To preserve ${\cal N}= 1$ one should further require that $i_{k} E 
=0$, which is a much stronger constraint and in our case this has 
solutions only in the limit $e^{\tilde K - K} = 0$.

At the same time, these considerations let us conclude that there is 
still the chance to find ${\cal N} = 1$ vacua in the decoupling limit 
where the hypermultiplets freeze.
In this limit one would end with ${\cal N} = 1$ supersymmetry 
preserved in the visible sector and ${\cal N} = 0$ in the hidden 
one \cite{susy2}.

\ 

We conclude this brief  analysis by recalling that the final form of 
the potential still respects the $SL(2,{\ZZ})$ invariance of the 
ten--dimensional action.
Since the reduction of the ten--dimensional dilaton/axion field to 
four dimensions is not straightforward, the action of such symmetry 
on the four--dimensional fields is non--trivial \cite{Halle}.
Anyway, we saw that the form of the potential one derives by direct 
compactification depends only on the ten--dimensional field $\eta$ 
and the rescaling needed to go to the Einstein frame.
Moreover, before this rescaling the potential was $SL(2,{\ZZ})$ 
invariant, once the appropriate  action on the charges was chosen 
(they must transform as vectors).
Since the quantity by which we rescaled it is 
proportional to the Calabi--Yau volume is also $SL(2,{\ZZ})$ 
invariant  the final potential preserves this symmetry.

\ 

\section*{Acknowledgments.}

\noindent I am glad to thank R. D'Auria and S. Ferrara for the many 
enlightening discussions and comments and for informing me on the 
related work in progress \cite{futuro}.
I also would like to thank J. Louis for informing me of related work 
in progress \cite{Louis} and A. Klemm and D. L\"ust for discussions.
This research is supported by the EC under RTN project HPRN-CT-2000-00131.

\vspace{2cm}

\providecommand{\href}[2]{#2}\begingroup\raggedright
\end{document}